# Purcell-enhanced single-photon emission from InAs/GaAs quantum dots coupled to broadband cylindrical nanocavities

Abhiroop Chellu[1,*], Subhajit Bej[2], Hanna Wahl[2], Hermann Kahle[1,3,7], Topi Uusitalo[1], Roosa Hytönen[1], Heikki Rekola[4], Jouko Lång[5], Eva Schöll[3,8], Lukas Hanschke[3,9], Patricia Kallert[3], Tobias Kipp[6], Christian Strelow[6], Marjukka Tuominen[5], Klaus D. Jöns[3], Petri Karvinen[4], Tapio Niemi[2], Mircea Guina[1], and Teemu Hakkarainen[1]

[1]Optoelectronics Research Centre, Physics Unit, Tampere University, Korkeakoulunkatu 3, 33720, Tampere, Finland

[2]Nanophotonics Group, Physics Unit, Tampere University, Korkeakoulunkatu 3, 33720, Tampere, Finland

[3]Institute for Photonic Quantum Systems (PhoQS), Center for Optoelectronics and Photonics Paderborn (CeOPP) and Department of Physics, Paderborn University, Warburger Straße 100, 33098, Paderborn, Germany

[4]Institute of Photonics, University of Eastern Finland, Yliopistonkatu 2, 80100, Joensuu, Finland

[5]Comptek Solutions Oy, Voimakatu 18, 20520, Turku, Finland

[6]Institute of Physical Chemistry, Universität Hamburg, Grindelalle 117, 20146, Hamburg, Germany

[7]Present address: Department of Physics and Astronomy, The University of New Mexico, 210 Yale Blvd NE, 87131, Albuquerque, United States of America

[8]Present address: Institute of Semiconductor and Solid State Physics, Johannes Kepler University, Altenberger Straße 69, 4040, Linz, Austria.

[9]Present address: Walter Schottky Institute, TUM School of Computation, Information and Technology and MCQST, Technical University of Munich, Am Coulombwall 4, 85748, Garching, Germany

*Email: abhiroop.chellu@tuni.fi

**Abstract**

On-chip emitters that can generate single and entangled photons are essential building blocks for developing photonic quantum information processing technologies in a scalable fashion. Semiconductor quantum dots (QDs) are attractive candidates that emit high-quality quantum states of light on demand, however at a rate limited by their spontaneous radiative lifetime. In this study, we utilize the Purcell effect to demonstrate up to a 38-fold enhancement in the emission rate of InAs QDs by coupling them to metal-clad GaAs nanopillars. These cavities, featuring a sub-wavelength mode volume of $4.5 \times 10^{-4}$ $(\lambda/n)^3$ and low quality factor of 62, enable Purcell-enhanced single-photon emission across a large bandwidth of 15 nm. The broadband nature of the cavity eliminates the need for implementing tuning mechanisms typically required to achieve QD-cavity resonance, thus relaxing fabrication constraints. Ultimately, this QD-cavity architecture represents a significant stride towards developing solid-state quantum emitters generating near-ideal single-photon states at GHz-level repetition rates.

**Main**

Epitaxially grown semiconductor quantum dots (QDs) have recently been at the forefront of technological developments in quantum key distribution[1–5] (QKD) and linear optical quantum computing[6]. The general performance of these applications relies on a triggered source generating non-classical light such as single photons and entangled photon pairs at a high repetition rate[7]. While QDs can generate state-of-the-art non-classical light states on demand[8–13], the overall obtainable count rate from them is limited by two factors. Firstly, the emission rate of a QD fabricated within a bulk semiconductor medium is limited by the intrinsic spontaneous radiative lifetime of its excitonic carriers. Secondly, only ~1 % of the emitted photons are reliably collected from within the high-index planar semiconductor layer(s) hosting the QD. Moreover, the isotropic emission profile of the QD constrains the collection efficiency and reduces the count rate. To this end, placing the QD within a photonic cavity largely addresses both problems. The modified local density of optical states (LDOS) within the cavity enhances the emission rate of the embedded QD via the Purcell effect[14,15], while the geometry of the cavity helps to efficiently out-couple the emitted photons.

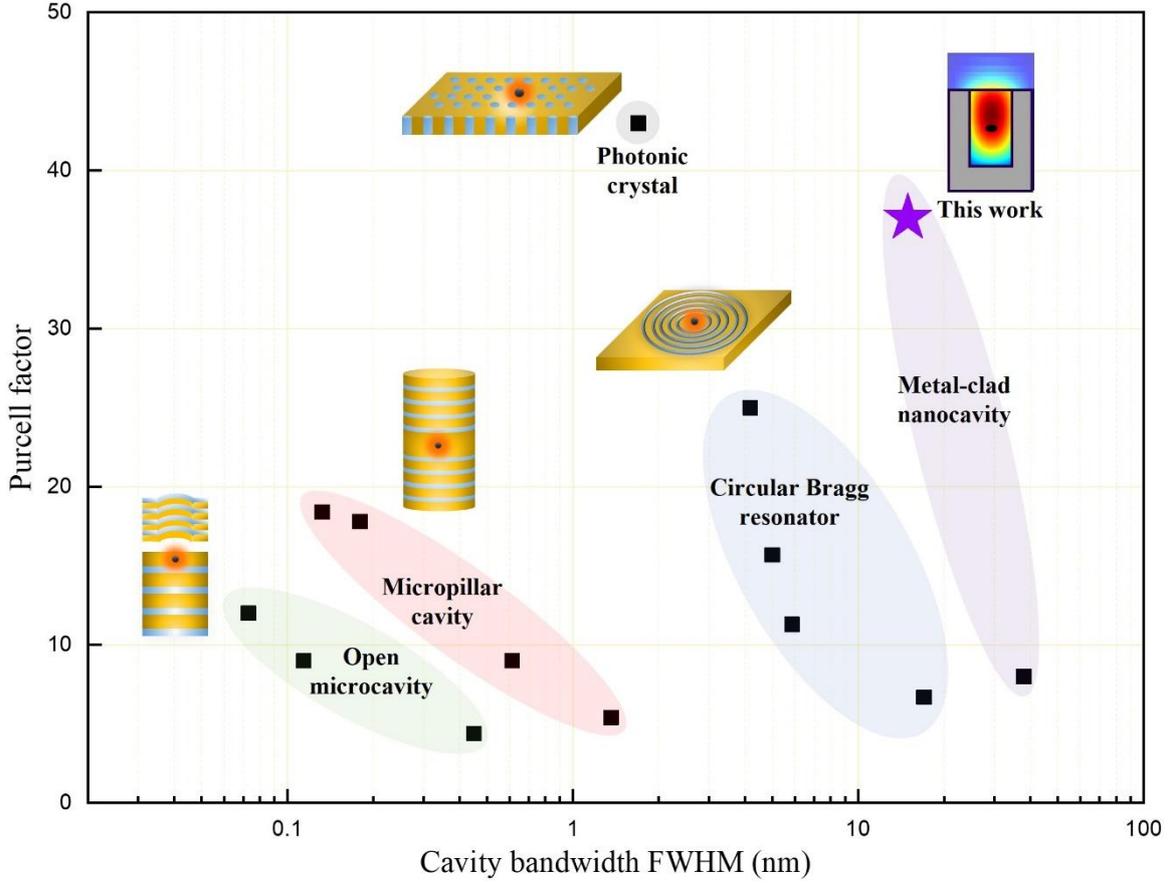

**Fig. 1 Purcell factor vs cavity bandwidth for different QD-cavity architectures.** Experimentally measured values of Purcell factor plotted as a function of the cavity bandwidth for III-V QDs embedded within various cavity architectures. References for the data points (from left to right within the zone highlighted for each cavity type): open microcavity[16–18], micropillar cavity[19–22], photonic crystal cavity[23], circular Bragg resonator[24,20,25,26], and metal-clad nanocavity[27].

Strategies for designing QD-cavity architectures often involve considering the important trade-off between achievable Purcell factor ($F_P$) and operational bandwidth of the cavity (as illustrated in **Fig. 1**). To understand the limitations imposed by this trade-off, we recall that the enhancement in emission rate, quantified by $F_P$, is proportional to the ratio between the quality factor ($Q$) and mode volume ($V$) of the cavity[15]:

$$F_P = \frac{3}{4\pi^2}\left(\frac{\lambda_c}{n_{eff}}\right)^3 \frac{Q}{V}. \tag{1}$$

Here, $\lambda_c$ is the resonance wavelength of the cavity and $n_{eff}$ is the effective refractive index within the cavity medium. An overwhelming majority of QD-cavity systems are designed with the motive to maximize $Q$ to ultimately increase $F_P$. Common photonic cavity architectures operating in the intermediate-$Q$ to high-$Q$ regimes (with $Q$ values typically within $10^3$–$10^4$) include open Fabry-Perot microcavities[16–18], micropillar-DBR cavities[8,19–22,28–30], and photonic crystal cavities[23,31–33] (**Fig. 1**). This approach, however, automatically limits the range of wavelengths over which Purcell enhancement is provided since $Q$ is inversely related to the bandwidth of the cavity.

In turn, a limited bandwidth imposes two practical limitations for integrating QDs within high-$Q$ cavities. The first limitation is caused by an inhomogeneity in QD dimensions introduced by the self-assembly based growth process. The underlying stochasticity governing the dimensions of a QD directly translates into a randomness in the average confinement potential it provides to excitonic carriers. This leads to a Gaussian spread of the central emission wavelength(s) of single QDs within an ensemble, which is typically orders of magnitude spectrally wider than the resonance bandwidth of a high-$Q$ cavity[34]. Consequently, the possibility of Purcell enhancement is relegated to a small fraction of QDs resonant with the narrow mode of the cavity. The second limitation arises from the inability of a high-$Q$ cavity to enhance more than one of many spectrally separate excitonic emissions originating from the same QD. For instance, QD-based sources relying on the biexciton (XX) – exciton (X) cascade process to generate entangled photon pairs benefit from the simultaneous enhancement of XX and X transitions, which are typically 1–3 nm spectrally apart[11–13,35,36]. To this end, single QDs are coupled with increasingly common low-$Q$ circular Bragg resonators (typically exhibiting $Q$ values of ~$10^2$) to facilitate broadband Purcell enhancement[20,25,26,37] (**Fig. 1**). Moreover, various tuning mechanisms[17,38–40] can be employed to match the QD emission to the cavity mode (and vice versa). However, implementing tuning mechanisms to achieve QD-cavity resonance inevitably raises the conundrum involving crucial trade-offs between the viable tuning range, compatibility, nanofabrication complexity, and the impact of detrimental effects, if any, on the optical properties of the QDs and the cavity alike.

In this work, we couple single InAs QDs to a relatively unexplored cavity architecture based on Ag-clad GaAs nanopillars that provide up to $F_P = 37.9 \pm 1.9$ (**Fig. 1**). Integrating dielectric GaAs nanopillars within a metallic surrounding allows for sub-wavelength confinement of light due to the large refractive index contrast along the Ag-GaAs interface(s)[27,41]. Consequently, the ultra-small $V$ of $4.5\times10^{-4}$ $(\lambda/n)^3$ and low $Q$ of 62, corresponding to a mode bandwidth of 15 nm, ensure excellent spatial and spectral overlap, respectively, between the embedded QD and the cavity mode. We employ a template-stripping based approach, which is termed as the *flip-strip* process, for experimentally realizing the cavities with a fabrication yield of 99.9 %. The probability of multi-photon emission of QDs embedded in these cavities is observed to reach as low as 0.7 % under pulsed quasi-resonant excitation. Remarkably enough, the ultra-short emission timescale of the fastest QDs causes a noticeable increase in their multi-photon emission due to a considerably high re-excitation probability[42]. This consequence is evident even when exciting the QDs with short (5 ps) excitation pulses.

**Designing the ultra-small mode volume cavities**

The design of the ultra-small $V$ cavities was supported by finite-difference-time-domain (FDTD) calculations carried out for a GaAs nanopillar (height, $h = 200$ nm, and diameter, $d = 100$ nm) surrounded on the sides and bottom by Ag (**Fig. 2a**). For this cavity geometry, the fundamental mode is confined to a central region within the nanopillar — over an estimated $V = 4.5\times10^{-4}$ $(\lambda/n)^3$ — due to the large refractive index contrast between GaAs and the surrounding Ag. An InAs QD is approximated as a point dipole with its dipole moment oriented parallel to the axis of the nanopillar. It lies in the vicinity of the optical mode field maximum when placed at the center of the nanopillar. Consequently, this allows the QD to experience an enhanced LDOS to favorably radiate into the cavity mode.

Next, the real and imaginary parts of the complex effective refractive index, Re($n_{eff}$) and Im($n_{eff}$), respectively, were simulated for an infinitely long Ag-clad cylindrical GaAs waveguide supporting the fundamental TE$_{11}$ mode (**Fig. 2b**). Their behavior as a function of the waveguide diameter, closely approximating $d$ in this case, reveals crucial information about the mode cut-off and losses within the medium. These simulations were performed using low-temperature values of refractive index values for Ag extracted from the surface plasmon propagation lengths[43]. Here, different refractive index values were used to simulate the behavior of the cavity mode in case of rough Ag (as-evaporated Ag film with RMS roughness of 4.5 nm) and smooth Ag (template-stripped Ag film with RMs roughness of 0.7 nm). Re($n_{eff}$) asymptotically decreases as the waveguide diameter becomes smaller, with the mode cut-off seen at $d \approx 95$ nm. It is important to note that the mode profile displayed in **Fig. 2a** results from operating in a regime above the mode cut-off. The behavior of Re($n_{eff}$) is practically the same regardless of whether rough or smooth Ag surrounds the medium. However, Im($n_{eff}$), representing losses within the medium, is observed to increase as the waveguide diameter becomes smaller. This behaviour is caused by the rapidly diminishing ability of the waveguide to confine light for dimensions approaching the mode cut-off. Noticeably, a rough Ag surrounding results in effectively higher Im($n_{eff}$) within the medium than in the case of smooth Ag owing to increased optical losses.

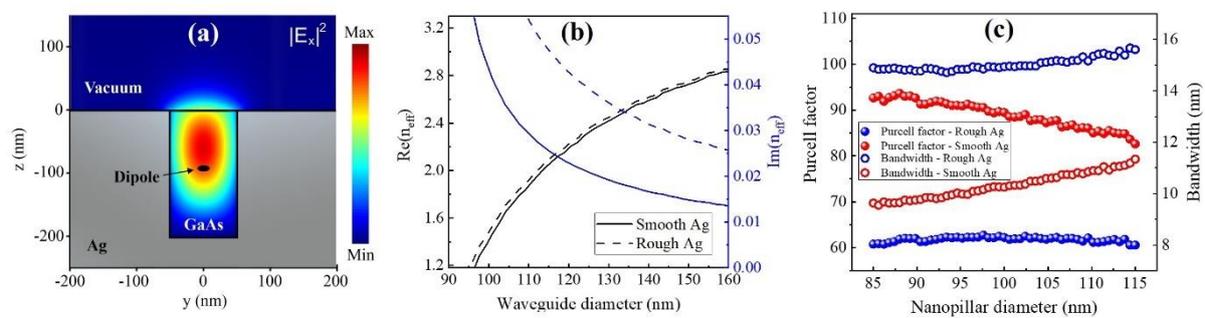

**Fig. 2 Key features of the simulated cavity mode. a,** FDTD simulation of the EM field in a GaAs nanopillar embedded within an Ag surrounding. A QD placed at the center of the nanopillar spatially overlaps well with the cavity mode, which is concentrated in the central region of the nanopillar volume. **b,** Behavior of the real and imaginary components of the complex effective refractive index within an infinitely long cylindrical GaAs waveguide surrounded by Ag. **c,** The dependence of the simulated $F_P$ and mode bandwidth on nanopillar diameter for the cases when rough and smooth Ag surround the cavity.

Following this, relevant performance metrics of the device such as $F_P$ and the cavity bandwidth were computed for nanopillars with $d$ = 85–115 nm and $h$ = 200 nm (**Fig. 2c**). The resonance wavelength of the cavity red-shifts as a function of increasing $d$ (**Fig. S1**), exhibiting behavior consistent for a medium with increasing Re($n_{eff}$). The full-width-at-half-maximum (FWHM) of the mode resonance, representing the cavity bandwidth, ranges from 10–12 nm as a function of increasing $d$ when the nanopillar is surrounded by smooth Ag. On the other hand, the bandwidth is ~15 nm regardless of $d$ when rough Ag surrounds the nanopillar. Likewise, the expected $F_P$ ranges from 90–80 as $d$ increases for a nanopillar with a smooth Ag surrounding, while remaining invariably ~60 with a rough Ag surrounding.

## *Flip-strip* nanofabrication process

An overview of the key steps involved in the *flip-strip* nanofabrication of devices is provided here. Further elaboration with relevant processing details for each step is provided in the Methods section.

- Epitaxial growth – The sample is grown with InAs QDs embedded in the middle of a 200 nm thick GaAs layer (**Figs. 3a** and **3b**).

- <u>Fabrication of nanopillars</u> – The GaAs layer is patterned and etched (**Figs. 3c, 3d** and **3i**) to result in seven distinct regions. Each region contains six separate 25×25 arrays of nanopillars of a particular diameter ($d$ = 85, 90, 95, 100, 105, 110, 115 nm).
- <u>GaInP under-etching</u> – The GaInP layer in the planar areas surrounding the nanopillars is etched away to weaken the mechanical attachment of the nanopillars to the growth substrate (**Figs. 3e** and **inset of 3i**).
- <u>Surface passivation</u> – Kontrox process (Comptek Solutions Oy) is used for conformal passivation of the GaAs nanopillars.
- <u>Metallization</u> – Conformal metal evaporation ensures that the nanopillars are uniformly coated and completely embedded within a 500 nm layer of Ag (**Fig. 3f**).
- <u>Flip-chip bonding</u> – A SiC carrier is bonded to either side of the sample after metallization (**Fig. 3g**).
- <u>Shear-stripping</u> – The SiC carrier bonded to the Ag film containing the embedded nanopillars is mechanically separated from the growth substrate using a lateral shear force (**Fig. 3h**).

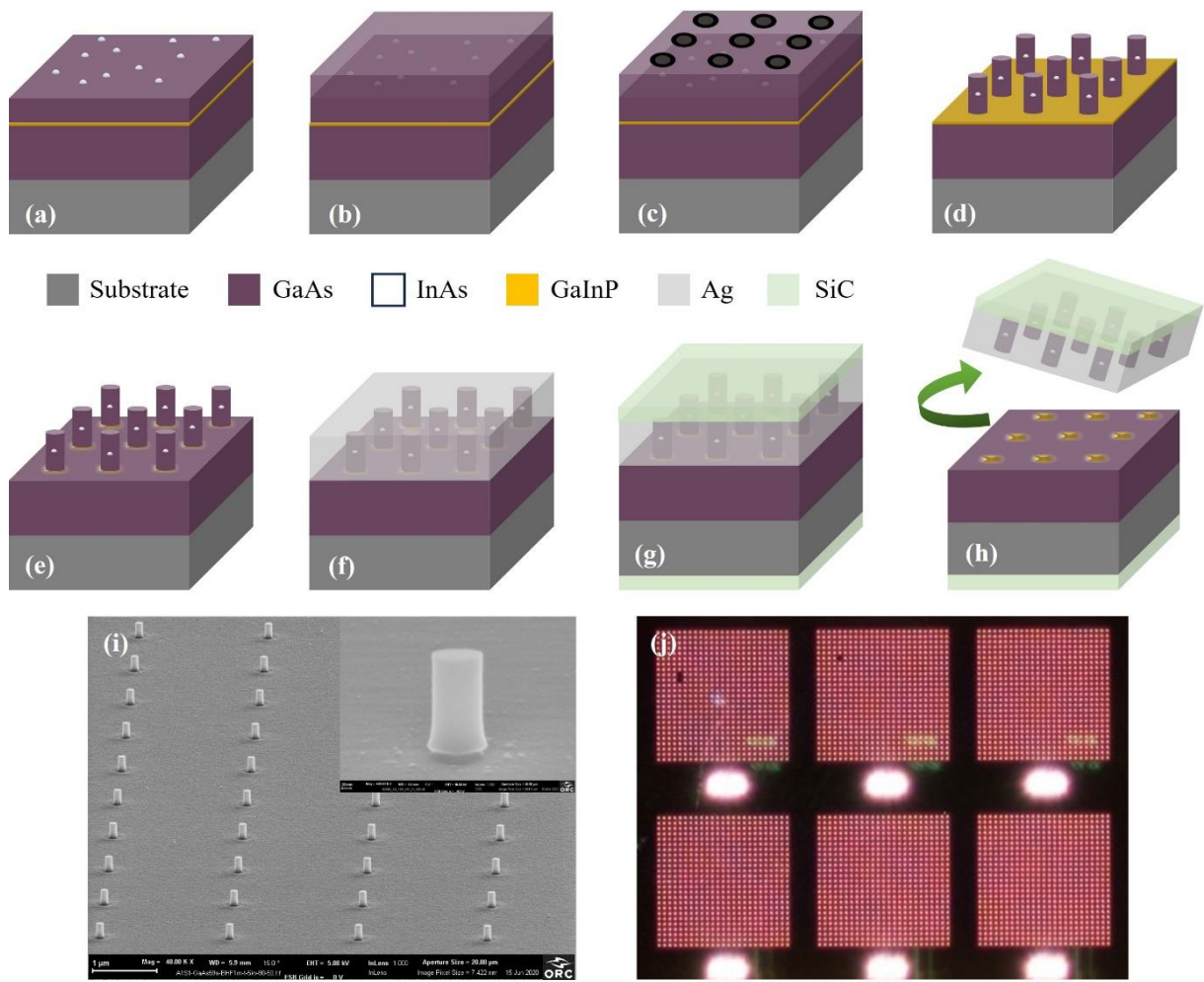

**Fig. 3 Illustration of the *flip-strip* nanofabrication process. a,** Self-assembly of InAs QDs on a GaAs surface. **b,** Capping the QDs with a GaAs layer. **c,** Patterning of nanopillar sites on the GaAs surface. **d,** Dry-etching resulting in GaAs nanopillars. **e,** Under-etching the GaInP layer surrounding the nanopillars. **f,** Metallization of nanopillars **g,** SiC carriers attached to metallized sample. **h,** Shear-stripping separates sample at the semiconductor-metal interface. **i,** SEM image showing an array of GaAs nanopillars after the dry-etching step. The inset shows a close-up of a nanopillar standing on a GaInP stub after the under-etching step. **j,** Dark-field optical microscope image shows the template-stripped Ag surface from the nanopillars|[100] region containing arrays of GaAs nanopillars embedded within the metal film.

The crux of the problem necessitating the *flip-strip* approach lies in the architecture of the sample after the metallization step. The fabricated structure at this stage contains QD-nanopillars completely embedded under 500 nm of Ag on the top while still attached to the underlying semiconductor layer stack in the bottom. This makes it impossible to optically access the QDs embedded within the nanopillars due to scattering of the excitation laser by the Ag surface from above and absorption by various semiconductor layers from below. Therefore, the underlying semiconductor layers along with the substrate need to be removed to expose the base of the nanopillars, which is the only metal-free interface through which the QDs can be optically accessed.

We resolve this by combining features from the standard template-stripping[44] and flip-chip nanofabrication processes. After the flip-chip bonding step, we perform shear-stripping to separate the structure favorably at the weakly adhesive semiconductor-metal interface. The applied shear force, complemented by the mechanically weak attachment of the nanopillars to the surface after the under-etching step, causes them to detach cleanly from the underlying semiconductor. Consequently, the stripped SiC carrier contains the Ag film hosting the nanopillars with their bases exposed and therefore allowing for optically addressing the embedded QDs. A similar approach employed to fabricate relatively larger metal-coated semiconductor structures achieved an overall process yield of 70 %[45]. A comparatively higher yield achieved through our *flip-strip* process is evident from **Fig. 3j** showing 99.9 % of the nanopillars transferred from the original semiconductor substrate to the new Ag host.

## QD-cavity emission characteristics

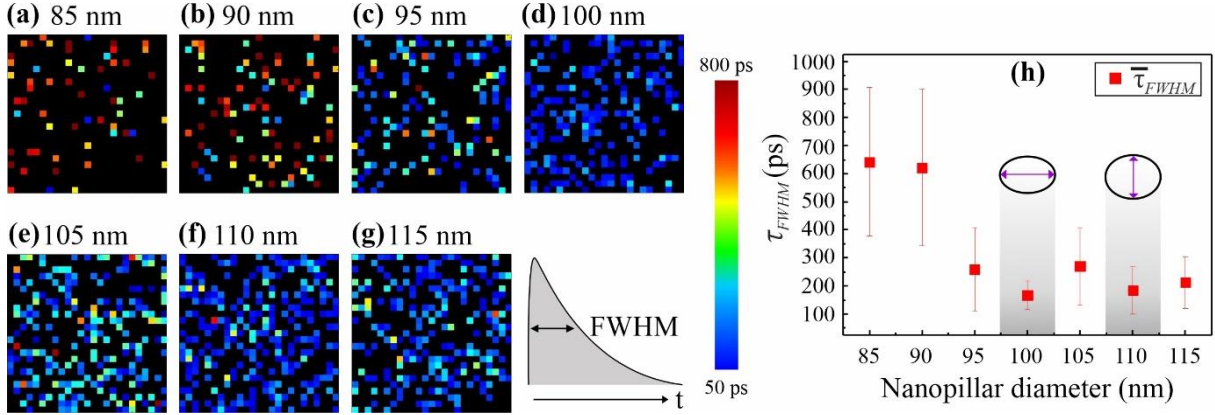

**Fig. 4 Mapping QD-nanopillar decay statistics. a - g,** TRPL data collected from every QD-nanopillar unit is plotted in a single 25×25 array for each distinct $d$ region. A single map contains 25 pixels in every row and column of the array, where each pixel represents the FWHM of the decay response of the corresponding QD. **h,** The mean value of QD lifetime, $\bar{\tau}_{FWHM}$ (with standard deviation), is plotted as a function of $d$. The shortest lifetimes (with 50 ps detector resolution) are observed for QDs within nanopillars|100 and nanopillars|110. Insets of this plot are used for suggesting the notion of a structural ellipticity present in the nanopillar shape that causes emission from QDs to preferentially couple to a particular polarization mode of the cavity. Here, QDs embedded within nanopillars|100 emit along the long axis while QDs embedded within nanopillars|110 emit along the short axis.

Bearing in mind the goal of identifying single QDs exhibiting the shortest Purcell-enhanced lifetimes among a vast number of QD-nanopillar units, we approached the task by first analyzing their time-resolved photoluminescence (TRPL) statistics. The FWHM of the decay curve, $\tau_{FWHM}$ (with 50 ps timing resolution determined by Si SPAD detectors), was extracted from the TRPL response of QD-nanopillars within a 25×25 array in each $d$ region. Every pixel in the resulting map represents the decay lifetime of a QD at the corresponding location (**Fig. 4a–4g**).

Evidently, a fewer number of QDs within the smaller nanopillars ($d$ = 85, 90 nm) are optically active, with this fraction noticeably higher for the larger nanopillars ($d$ = 95, 100, 105, 110, 115 nm). This is expected because the smaller cavities have statistically lower chance of hosting a QD, a constraint that relaxes as $d$ increases. Moreover, QDs within the smaller cavities are more likely to be laterally closer to GaAs surface states in the nanopillar sidewalls, making them more prone to surface effects and potentially rendering them optically inactive[46–48]. On the other hand, the requirements for the lateral positioning of QDs within the larger nanopillars are more relaxed, which ultimately increases their likelihood of being optically active. Additionally, the observed distribution of optically active QDs can be considered a direct consequence of the larger nanopillars operating above the mode cut-off (**Fig. 2b**).

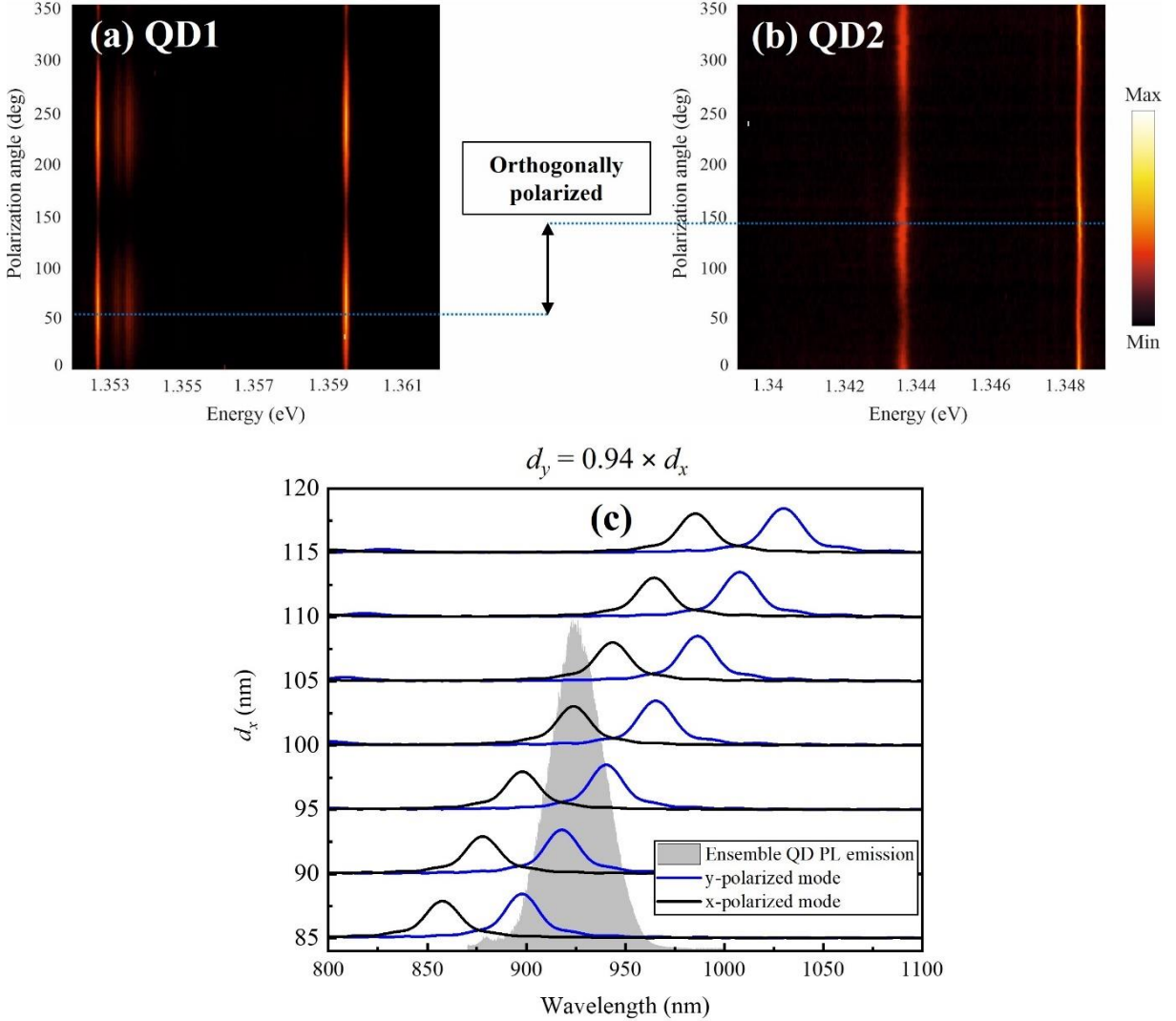

**Fig. 5 Single-QD polarization-resolved PL and simulation of modes in an elliptical nanopillar.** Polarization-resolved PL spectrum of **a,** QD1 embedded in a nanopillar|100 and **b,** QD2 embedded in a nanopillar|110. The comparison of variation in the emission intensity of excitonic peaks reveals a 90° phase-shifted polarization response between the two QDs. **c,** A simulation of the cavity mode(s) in an elliptical nanopillar where the diameter along y-direction is 94 % of the diameter along x-direction is shown. The mode is split into orthogonal x- and y-polarized components with a spectral separation of 41 nm between them. The y-polarized mode in nanopillars|100 and x-polarized mode in nanopillars|110 are expected to support the QD emission. The spectral mismatch between the ensemble QD emission peak and the supporting cavity modes can arise from using imprecise low-temperature refractive index values in the simulations.

The mean QD lifetime, $\bar{\tau}_{FWHM}$, which is calculated as the mean of $\tau_{FWHM}$ from QD-nanopillars within a 25×25 array, is clearly dependent on $d$ (**Fig. 4h**). The QDs within the smaller nanopillars exhibit $\bar{\tau}_{FWHM}$ = 600–650 ps while QDs within the larger nanopillars decay faster with $\bar{\tau}_{FWHM}$ = 150–250 ps. The QDs within nanopillars with $d$ = 100 nm (nanopillars|100) exhibit the smallest $\bar{\tau}_{FWHM}$ because their mode resonance (**Fig. S1**) matches well with the QD ensemble emission wavelength (**Fig. S2b**).

Incidentally, the QDs within nanopillars|[110] exhibit the second-smallest values of $\bar{\tau}_{FWHM}$. This trend in the statistics gives rise to the notion of an inherent structural ellipticity of the nanopillars creating an asymmetric cavity mode supporting two orthogonal polarization components. This assumption is verified by analyzing the polarization-resolved PL spectra of single QDs. The maximum emission intensity of excitonic transitions in QD1 within a nanopillar|[100] occurs at a relative polarization angle of 55°. Whereas the emission maximum from QD2 within a nanopillar|[110] is seen at a relative polarization angle of 145° (**Figs. 5a** and **5b**). The nature of the cavity mode simulated for an elliptical nanopillar where the diameter along y-direction, $d_y$, is 94 % of the diameter along x-direction, $d_x$, provides further validation for these observations (**Fig. 5c**). The cavity mode is split into x- and y-polarized components with a spectral difference of 41 nm between them. Based on the relative cavity dimensions, QDs in nanopillars|[100] couple to the y-polarized mode and emit along the elongated side while QDs within nanopillars|[110] couple to the x-polarized mode and emit along the shortened side. Deviations from this ideal scenario are expected when the QD is displaced from the center of the nanopillar. The Purcell enhancement experienced by off-axis QDs can decrease as a function of their offset, while their polarization response can experience shifts proportional to their radial displacement[41] (**Fig. S5**).

**Emission characteristics of single QDs**

The ratio between the emission lifetime of QDs in the absence of a cavity, $\tau_0$ = 856.4 ± 43.5 ps (**Fig. 6a**), and the emission lifetime of QDs within nanopillars, $\tau$, represents a quantitative estimate of the Purcell enhancement, such that $F_P = \tau_0/\tau$. An emission lifetime of $\tau$ = 22.6 ± 0.7 ps was measured for an excitonic transition in QD1 under quasi-resonant p-shell excitation using 5 ps pulses (**Fig. 6a**). This Purcell-shortened lifetime corresponds to a significant $F_P$ of 37.9 ± 1.9. The linewidth of the emission peak of QD1 shown in the inset of **Fig. 6a** is determined to be 36.6 µeV (**Fig. S8**), which is larger than the Fourier-limited linewidth by a factor of only 1.25. The second-order autocorrelation function, $g^{(2)}(0)$ = 0.1188 ± 0.0185 (**Fig. 6c**), measured for the same transition suggests a low multi-photon emission probability from QD1, thus affirming its quality as a single-photon emitter.

Interestingly, QDs which are characterized with $\tau$ comparable to the duration of the excitation pulse are observed to be affected by a significant probability of re-excitation. In the event of re-excitation, a QD can exhibit two separate single-photon emission events within the duration of a single excitation pulse[42,49,50]. Such an occurrence inadvertently increases the measured $g^{(2)}(0)$ of the source. As shown in **Fig. 6b**, a considerably high re-excitation probability of 5–10 % is estimated for QDs characterized with $\tau$ <50 ps. It is important to note here that the re-excitation probability also depends on the time taken by excited carriers to relax from the p-shell to s-shell of the QD. By comparing the simulated re-excitation probabilities with the measured $g^{(2)}(0)$, it is possible to determine a p-s relaxation jitter of approximately 5–10 ps in these QDs. Contrastingly, QDs with longer lifetimes are observed to experience a relatively lower probability of re-excitation. For instance, QD3 characterized with a $\tau$ = 223.3 ± 1.9 ps (**Fig. 6a**) and a $g^{(2)}(0)$ = 0.0078 ± 0.0013 (**Fig. 6d**) is seen to experience a negligibly low re-excitation probability. Notably, the multi-photon emission probability observed for QD3 is up to an order of magnitude lower than other reports for quasi-resonantly excited In(Ga)As/GaAs QDs[51–53]. Nevertheless, these findings related to re-excitation probability provide an incentive to adopt resonant excitation schemes[54] for further suppressing the multi-photon emission probability of these QDs.

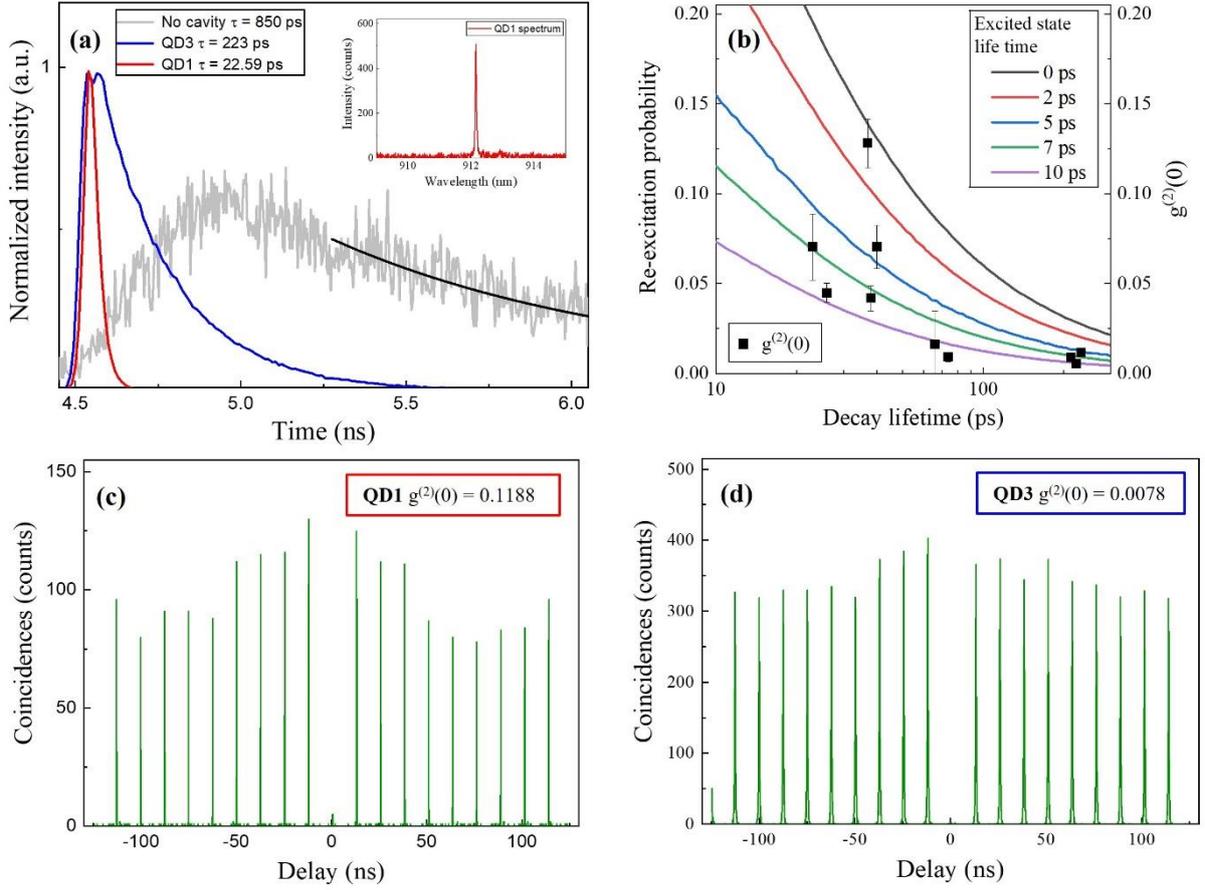

**Fig. 6 Purcell-enhanced lifetime and multi-photon emission probability of single QDs. a,** Purcell-enhanced decay curves of quasi-resonantly excited QD1 (red line) with $\tau$ = 22.6 ± 0.7 ps and QD3 (blue line) with $\tau$ = 223.3 ± 1.9 ps. A decay response of $\tau_0$ = 856.4 ± 43.5 ps is obtained by exciting an ensemble of QDs via the wetting layer in the absence of a cavity (gray line). The data is fitted with a single-exponential function to guide the eye (black line). The inset shows the spectrum of quasi-resonantly excited QD1 with a single excitonic transition around 912 nm. **b,** The simulated re-excitation probability expected for different excitation pulse durations is shown as a function of decay lifetime of QDs. It is seen that the QDs with $\tau$ <50 ps suffer from a higher re-excitation probability that degrades their experimentally measured $g^{(2)}(0)$ values. **c,** Measured single-photon autocorrelation for QD1. **d,** Measured single-photon autocorrelation for QD3.

## Discussions

We have demonstrated Purcell-enhanced single-photon emission from InAs QDs embedded within Ag-clad GaAs nanopillars. The sub-wavelength mode-volume in these cavities allows for single-mode operation while their low-$Q$ nature provides the significant advantage of achieving QD-cavity resonance without employing complicated tuning mechanisms. These attractive cavity-related features have allowed for achieving $F_P$ as high as 38, which compares well with the state-of-the-art for III-V QD-cavity systems. At this stage, it is important to gain insight into another important performance metric, namely the source efficiency, of this QD-cavity system. This can be calculated from the total efficiency with which the embedded QD emits into the upper half space of the nanopillar, $\eta(\pi)$. In this scenario, the photons emitted by the QD are lost via three different mechanisms: (i) absorption losses experienced by the cavity mode, (ii) losses via direct coupling to surface plasmons, and (iii) losses resulting from collection inefficiencies.

The first type of loss is caused by the penetration of the cavity mode beyond the physical confines of the GaAs nanopillar and into the surrounding Ag medium. This results in the non-radiative dissipation of the emitted optical energy via absorption in the Ag layer. Consequently, this loss mechanism is largely governed by the imaginary part of the complex permittivity of Ag, denoted as Im($\varepsilon_{Ag}$), which is determined by the roughness at the Ag/GaAs interface as well as by the general material quality of the evaporated Ag. The second mechanism manifesting as losses is caused by the direct coupling of emitted photons to surface plasmons propagating along the planar Ag surface. The probability of direct photon-plasmon coupling in a nanopillar|[100] is estimated to be approximately 5 %, which constitutes only a small fraction of the total loss[41]. It is important to note here that this coupling mechanism does not contribute to the fast decay rates observed for QDs embedded within nanopillars (**Fig. 4**). This is because relatively slower decay rates are observed for QDs in the smallest cavities (i.e. nanopillars|[85] and nanopillars|[90]), where the coupling rate to interface plasmons should be strongest. Finally, the third loss mechanism affecting the overall source efficiency arises from an inefficiency in extracting the emitted photons by the objective lens. The photon extraction efficiency, $\eta(0.81)$, is determined as the total power emitted into the cone that can be harvested by an objective lens with N.A. = 0.81. The cavity characterized with nanopillar dimensions of $d = 100$ nm and $h = 200$ nm is referred to as **Geometry 1** for the rest of this section. For this structure, we get $\eta(\pi)$ of 42 % and 29 % for smooth and rough Ag, respectively, while the corresponding values of $\eta(0.81)$ are 24 % and 17 %, respectively (**Table 1**). As shown in **Fig. 7**, the emission from **Geometry 1** diverges into large angles in the direction parallel to the emission polarization.

**Table 1.** Performance metrics of the QD-cavity source estimated from FDTD simulations.

|  | $\lambda_c$ (nm) | $Q$ | $F_P$ | $\eta(\pi)$ | $\eta(0.81)$ |
|---|---|---|---|---|---|
| **Rough Ag** | | | | | |
| Geometry 1 | 953 | 45 | 62 | 29 % | 17 % |
| Geometry 2a | 1009 | 23 | 34 | 58 % | 34 % |
| Geometry 2b | 944 | 41 | 51 | 42 % | 25 % |
| Geometry 2c | 950 | 41 | 52 | 46 % | 28 % |
| **Smooth Ag** | | | | | |
| Geometry 1 | 948 | 64 | 88 | 42 % | 24 % |
| Geometry 2a | 1004 | 26 | 39 | 67 % | 39 % |
| Geometry 2b | 940 | 53 | 67 | 55 % | 32 % |
| Geometry 2c | 945 | 52 | 66 | 59 % | 37 % |

A stepwise strategy involving geometric changes to the original cavity structure (**Geometry 1**) is presented next. This allows for exploring the possibility of developing a modified cavity design that experiences reduced losses while maintaining high $F_P$. The absorption losses can be minimized by increasing the nanopillar diameter, thus reducing the penetration of the cavity mode into the surrounding Ag. A reduction of the nanopillar height is required to compensate for the resulting increase in $n_{eff}$ of the cavity mode to maintain the cavity resonance at the same wavelength range as in the original structure. Thus, changing the cavity dimensions to $d = 152$ nm and $h = 93$ nm (**Geometry 2a** shown in **Fig. 7**) nearly doubles both $\eta(\pi)$ and $\eta(0.81)$ compared to **Geometry 1**, while $F_P$ and $Q$ are reduced to half due to the reduced reflectivity at the top surface. In the next step, depicted as **Geometry 2b** in **Fig. 7**, the cavity is recessed by 38 nm into the metal layer, thereby increasing the reflectivity at the top surface. Consequently, $F_P$ and $Q$ in this modified structure match the values witnessed in the original structure albeit with a slight loss in efficiency in comparison with **Geometry 2a**. Finally, rounded edges are introduced at the point where the recess meets the planar surface (**Geometry 2c** in **Fig. 7**). This causes a funnelling effect for photons originating from the cavity mode to couple into free space, thus resulting in a symmetric far-field pattern that ultimately improves the extraction efficiency.

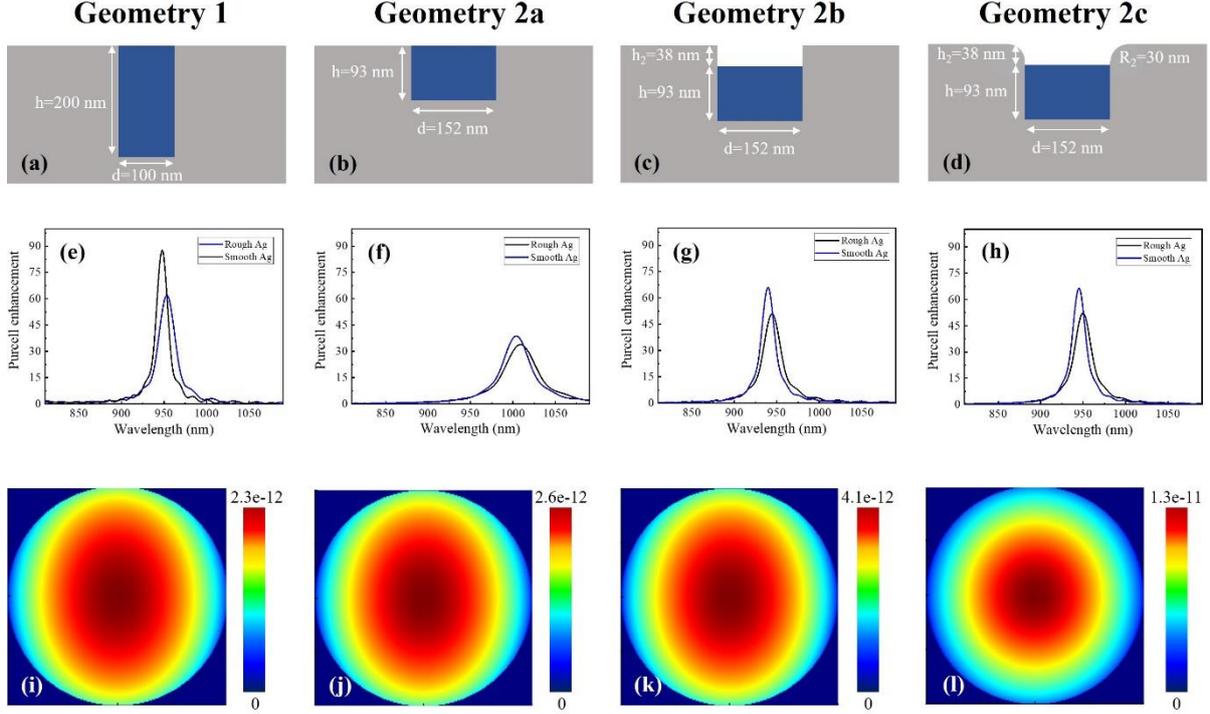

**Fig. 7 Simulation of performance metrics in modified cavity designs. a–d,** Series of schematics showing a stepwise modification of the cavity architecture from the original Geometry 1 to Geometries 2a–c. **e–f,** Simulation of Purcell enhancement **i–l,** and far-field mode profile of the corresponding geometries.

It is evident that the source efficiency of the cavity is strongly correlated with $F_P$ and $Q$ through the properties of the Ag surrounding the nanopillar. Therefore, the overall performance of the cavity can be significantly enhanced by simply improving the Ag material quality and reducing the roughness of the Ag/GaAs interface. To give some perspective of the achievable improvement, a value of $Im(\varepsilon_{Ag}) = 0.5$ at a wavelength of 1440 nm has been experimentally demonstrated for an Ag-clad InP nanolaser[55]. In comparison, the dispersion data used in our work for simulating cavity behaviour with smooth Ag surrounding the nanopillar amounts to $Im(\varepsilon_{Ag}) = 1.8$ at the same wavelength[43]. Thus, the performance metrics can be improved beyond the simulation results shown in **Table 1** by improving the quality of the Ag layer through careful optimization of the nanofabrication process. For instance, gentle wet-etching can be employed to polish the dry-etched sidewalls of the nanopillars in addition to further optimizing the evaporation parameters for forming the Ag layer.

The experimentally observed value of $F_P = 37.9 \pm 1.9$ is lower than the estimated value of 62 in the simulations for **Geometry 1**. Therefore, it is reasonable to conclude that the actual efficiency of the investigated devices is lower than what has been estimated in the simulations. The experimentally observed $F_P$ can be further limited by the misalignment of the QD from the center of the nanopillar (**Fig. S5**). To this end, employing QD-cavity localization techniques[56] would maximize the Purcell enhancement while eliminating the displacement-induced polarization effects (**Fig. S5b**), thus allowing this cavity design to support the generation of highly entangled photon pairs once the ellipticity is eliminated. The large mode bandwidth of 15 nm in this low-$Q$ cavity is sufficient for the simultaneous Purcell enhancement of XX-X transitions in state-of-the-art entangled photon sources such as, In(Ga)As/GaAs QDs[11,13], GaAs/AlGaAs QDs[12], and InAs/InP QDs[57]. Moreover, this bandwidth is sufficient for enhancing XX-X transitions with larger spectral separations typically seen in high-temperature QDs[58], thus enabling the realization of solid-state room-temperature entangled photon sources. Furthermore, a bandwidth of this magnitude opens avenues to study cooperative radiative phenomena like superradiance[59]. This can be realized, for instance, by coupling multiple QDs fabricated via correlated epitaxial growth of stacked QDs[60] to a broadband cavity mode.

On a more general note, the demonstrated *flip-strip* process is a key technological aspect underpinning the ability to fabricate the low-$Q$ cavities. This is a high-yield process involving a controllable chip-scale transfer of GaAs nanopillars containing QDs from their original substrate onto a new Ag host. The versatility of this technique, stemming from the generally weak adhesion between dielectrics and noble metals, can be extended to realize quantum emitters in other prominent hybrid platforms like plasmonic resonators[61], plasmonic lasers[62], or hybrid nanowires.

The benefits of a high Purcell enhancement factor are directly linked to the possibility of improving the performance metrics of QD sources. For instance, the demonstrated $F_P$ close to 38 corresponds to near-unity single-mode coupling efficiency, $\beta = F_P/(1+ F_P) \approx 97.4$ %. Moreover, the potential to reach gigahertz-level repetition rates with a QD source that exhibits a multi-photon emission probability as low as 0.7 % represents a marked advancement in the development of solid-state single-photon sources. With further development, these sources can be adopted in the future for resource-intensive applications like device-independent QKD and measurement-based quantum computing with photonic cluster states[7].

## Methods

**Design and simulation of the ultra-small V cavities** The cavity mode shown in **Fig. 2a** was simulated by FDTD method with a total-field scattered-field source while the behavior of $n_{eff}$ of the waveguide mode reported in **Fig. 2b** was simulated with an EM mode solver. The $F_P$ in **Fig. 2c** was simulated by FDTD by assessing the increase of dipole source power within the cavity with respect to the same dipole in a homogeneous dielectric environment. Experimentally determined low-temperature refractive index values of GaAs[63] and Ag[43] were used in all the simulations. In order to accurately estimate the cavity-related parameters, the simulations were carried out using two different refractive index values of Ag depending on the roughness of the Ag interface around the GaAs nanopillar. The necessity for doing so arises from the realistic nature of the critical interface between the GaAs nanopillar sidewalls and Ag which affects the cavity losses. To elaborate, the sidewall surface of an etched nanopillar is typically rougher than an as-grown semiconductor surface due to nanofabrication-induced roughness. Thus, the interface roughness of an Ag film directly evaporated on an etched surface (rough Ag) is typically higher than that of an Ag film directly evaporated on a flat epitaxially grown semiconductor surface (smooth Ag). This difference in interface roughness results in different refractive index values for the two Ag variants which ultimately affects the magnitude of optical losses experienced by the cavity mode.

## Nanofabrication of devices

**Epitaxial growth** Epitaxial growth of the sample(s) was carried out on (100)-oriented SI-GaAs substrates in a solid-source molecular beam epitaxy (MBE) system. The InAs QDs were grown in the Stranski-Krastanov growth mode via strain-induced self-assembly on a 100 nm GaAs matrix at 530 °C (**Fig. S2a**). Following this, indium-flushing was carried out to tune the emission wavelength of the QD ensemble to 925 nm. The FWHM of the emission peak, estimated to be 27 nm, is characteristic of the QDs' size homogeneity (**Fig. S2b**). The QD density was optimized to $2\times10^{10}$ cm$^{-2}$. This was done to maximize the probability for a nanopillar|$^{100}$ to host a single QD within 10 nm laterally from its center and minimize the probability of hosting multiple QDs. Finally, the QDs were overgrown with a 100 nm GaAs capping layer. The complete layer structure of the sample(s) investigated in the optical experiments is presented in (**Fig. S2c**).

**Fabrication of nanopillars** A 5 nm layer of Ti was evaporated on the GaAs capping layer to improve adhesion between the GaAs surface and the spin-coated silsesquioxane-based negative resist SX AR-N 8200 (Medusa 82). Electron beam lithography was used to pattern the resist with nanopillar masks with the nominal diameter of the patterns ranging from 85 nm to 115 nm in 5 nm incremental steps. Developing the patterned resist removed the unexposed resist material, thus leaving behind only the polymerized resist in the exposed areas acting as an etch mask for the subsequent dry-etching step. The mask pattern was first etched through the Ti adhesion layer and then through the GaAs-QD layer until the GaInP etch-stop using $Cl_2/N_2$ plasma in an inductively coupled plasma – reactive ion etching system. This formed GaAs nanopillars standing on the surface of the underlying GaInP etch-stop layer. The Ti-Medusa etch mask remaining on top of the nanopillars was then completely removed by dipping the sample in a BHF solution for 30 s.

**GaInP under-etching** By dipping the sample in a fresh 12 M HCl solution for 1 s, most of the GaInP lying in the planar areas surrounding the nanopillars was etched away. Only a small GaInP stub, designed to remain beneath each nanopillar, mechanically held the nanopillars in their place on the surface (**Fig. 3e** and **inset of 3i**).

**Surface passivation** Surface passivation largely negates the degrading effects of GaAs surface states on the optical properties of nearby QDs. To this end, the commercially available Kontrox process was employed to provide a conformal coating of an atomically precise ordered oxide layer on the nanopillar sidewalls.

**Metallization** The passivated sample was metallized in the vacuum chamber of an electron-beam metal evaporator at room temperature. The sample was mounted in a downward-facing 45° angle with respect to the Ag source. Rotating the sample at a constant speed ensured conformal and uniform coverage of Ag on the nanopillar sidewalls.

**Flip-chip bonding** The top and bottom surfaces of the metallized sample were bonded to a SiC carrier using a commercially available cryo-compatible epoxy. The epoxy was thermally cured overnight at 60 °C.

**Shear-stripping** A mechanical shear force accompanied by a gentle upward torque was applied on one corner of the SiC carrier bonded to the Ag film. The stronger adhesion between Ag and the SiC carrier than the adhesion between Ag and the underlying semiconductor surface caused the applied force to separate the sample at the semiconductor-metal interface.

## Optical characterization

**Mapping QD-nanopillars emission characteristics** The sample was mounted on a piezoelectric stage inside a closed-cycle cryostat and cooled to 8 K. A Ti:sapphire laser (Coherent Mira Optima F) generating pulses with a nominal pulse width of 150 fs and tuneable between 700–1000 nm was used as the excitation source. The laser beam, tuned to 880 nm, was guided through a polarizer and a half-wave plate (HWP) to adjust its polarization, a 900 nm short-pass filter and finally through a beam splitter (BS) with 90 % transmission and 10 % reflection. The reflected beam was guided to a two-axis Galvo-mirror scanner reflecting the light through a telecentric lens system into the entrance aperture of a microscope objective (N.A. = 0.85) placed inside the cryostat. The objective focused the laser beam to a diffraction-limited spot of ~600 nm in diameter and collected the PL signal from the sample. The collected PL signal, after passing through the telecentric lens system, the Galvo-mirror scanner, and the BS, was filtered by a 900 nm long-pass filter and a rotatable HWP placed in front of a polarizer. The filtered PL signal was then detected and resolved either spectrally or temporally. The PL signal was passed through a 500 mm spectrograph (Princeton Instruments Acton SP2500i) equipped with a 300 lines/mm grating at 750 nm blaze wavelength. A cooled back-illuminated, deep-depletion CCD camera (Princeton Instruments, Pixis 400BR) at the end of the spectrograph detected the spectrally resolved signal. Time-resolved measurements were performed by time-correlated single-photon counting. A silicon avalanche photodiode (Micro Photonic Devices, PDM Series) connected to a photon-counting electronics module (Picoquant PicoHarp300) detected the signal with 50 ps temporal resolution (estimated from the FWHM of the measured transient spread of the reflected laser light).

A rough localization of the nanopillar arrays was done by manually driving the stepper piezos of the stage inside the cryostat to an approximate position and subsequently rapidly scanning the laser focus over the sample with the Galvo-mirror scanner while monitoring the back-reflected signal. Here, scanning and time-integrated photon counting was done via self-written LabView software running a real-time computing and controlling system (Jäger Computer-gesteuerte Messtechnik GmbH, ADwin-Gold II). After the rough localization, the individual nanopillars within an array were scanned in the same way with an additional 900 nm long-pass filter to create a precise PL map QD-nanopillar units as exemplarily shown in **Fig. S3a**.

For the time- and spectrally resolved characterization of all nanopillars inside an array, we developed a procedure based on self-written MATLAB scripts that allowed us to collect signals only from appropriate nanopillars and thus drastically reduced the measurement time. We first determined the initial positions of the nanopillars (**Fig. S3b**) using the precisely plotted PL map mentioned above as a guide (see **Fig. S3a**). Next, we obtained the measured PL spectra by automatically addressing the determined positions of all nanopillars. Due to the sharp and bright spectral lines emitted by the QDs inside the nanopillars, the quality of the spectra could be estimated within an integration time of ~1 second. An overview of the spectra collected from the nanopillars exhibiting PL emission are shown exemplarily for one array in **Fig. S3d**. Following this, we created a list of coordinates from appropriate nanopillars. Here, we manually inspected a panel of all spectra from the array of nanopillars and considered only the coordinates with PL intensities above a certain threshold. Finally, the decay curves of the appropriate pillars were measured automatically. An overview of these decay curves is shown exemplarily for one array in **Fig. S3c**. The last step was the most time-consuming one of this procedure since relatively many more photons in comparison to the PL spectra were needed to obtain an appropriate decay curve. To further reduce the measurement time, the integration time of every decay curve was chosen dynamically for every position until a certain number of photons were collected. Thus, the integration time for nanopillars with bright PL emission could be reduced which allowed for longer integration time for the nanopillars exhibiting less intense emission. By following this procedure, we obtained a coarse spectrum from every nanopillar position in an array and decay curves for the nanopillars showing an appropriate PL signal.

**Single-QD polarization-resolved PL** The sample was mounted in a closed-cycle cryostat and cooled to 6 K. A diode laser operating in continuous-wave mode and emitting at 850 nm was used for non-resonantly exciting the QDs via the wetting layer. The excitation laser was focused to a spot size of ~1 μm using an objective (N.A. = 0.8) in order to selectively excite single QD-nanopillars. The same objective was used for collecting the PL emission from the QD. The PL signal then passed through a sequence of HWP, polarizing BS and 905 nm long-pass filter before entering a 750 mm spectrometer. An 1800 lines/mm grating split the incident PL signal with a spectral resolution of ~15 μeV, which was then detected using a water-cooled CCD detector array. By rotating the HWP using an automated rotation mount, we collected polarization-dependent PL maps for single QDs like the ones shown in **Figs. 5a** and **5b**.

**Single-QD and ensemble-QD TRPL** TRPL experiments were performed to estimate $\tau$ from single QD-nanopillars in a sample mounted in a closed-cycle cryostat and cooled to 5 K. A Ti:saphhire laser (Coherent Mira) generating 5 ps long pulses at a repetition rate of 80 MHz was used as the excitation laser. The wavelength of these pulses was tuned to excite carriers directly into the p-shell of a particular QD. A grating-based filter was employed to further spectrally isolate emission from the selected radiative transition of the QD. The PL emission was then directed to a superconducting nanowire single photon detector (SNSPD) characterized with a dark count rate of <1 counts/s and timing jitter of 20 ps. The radiative lifetime of the QD was analyzed from the histogram of arrival time of the emitted photons on the SNSPD. To account for the finite temporal resolution of the setup, the radiative lifetime of the QD was obtained by fitting the data with the convolution of a Gaussian function with a single exponential decay.

The decay lifetime of QDs in the absence of a cavity, $\tau_0$, was estimated from a sample containing QDs situated within a 200 nm thick planar GaAs layer and grown under the same conditions as the other sample(s). The sample was cooled to 6 K. The ensemble of QDs was excited via their wetting layer using a diode laser generating 100 ps long pulses with 80 MHz repetition rate at 850 nm and detected using a Si SPAD (50 ps resolution). The obtained decay curve was deconvoluted with the instrument response function and fitted to a single-exponential decay in order to estimate $\tau_0$.

**Single-photon autocorrelation measurements** The probability of multi-photon emission from single QDs was determined with a Hanbury Brown and Twiss type experimental setup. Here, the filtered PL signal was split by a 50:50 BS and detected by two SNSPDs (same SNSPD specifications as in single-QD TRPL experiments). A coincidence histogram was obtained by correlating the single-photon detection events occurring at the two SNSPDs. The value of the autocorrelation function at zero time delay position in the histogram, $g^{(2)}(0)$, of a single QD was then calculated from the ratio between the average number of coincidences occurring at non-zero time delay and number of coincidences at zero time delay. In order to account for blinking in the QD emission, the measured blinking data was fitted with a model, $y_0 = A*\exp(-abs(x-x_0)/t)$, that allowed for determining the on/off ratio, $A/y_0$, for the QDs. The $g^{(2)}(0)$ values were then corrected for blinking by multiplying with the on/off ratio.

**Estimation of re-excitation probability**

The re-excitation probability presented in **Fig. 6b** was calculated with a numerical probability simulation using a Gaussian distribution for the excitation source and a single-exponential decay for the emission. The Mersenne-Twister random number generator was used to simulate the probabilistic nature of the process. The time of the first excitation event was calculated from this probability simulation. The time of emission was calculated from the probability simulation starting from the moment when the first excitation took place. The probability of re-excitation was then calculated by integrating the tail of the Gaussian pulse from the moment of first emission to infinity. This process was repeated 1000000 times to produce a simulated estimation of re-excitation probability expected when using Gaussian excitation pulses of 5 ps long. A relaxation time jitter of 0, 2, 5, 7, and 10 ps resulting from the time taken for the carrier(s) to move from the excited state (p-shell) to the ground state is considered here.

**Data availability**

All data supporting the findings presented in this study are available within the main manuscript and the Supplementary Information. The raw data files are available from the corresponding author upon reasonable request.

**Code availability**

The codes used in this work are available from the corresponding author upon reasonable request.

**Acknowledgements**

The authors acknowledge Jari Lyytikäinen for his support in MBE-related work. Turkka Salminen is acknowledged for his assistance with focused ion beam milling. A.C., R.H., M.G., and T.H. acknowledge funding support from Research Council of Finland's projects NanoLight and QuantSi as well as Business Finland's co-innovation project QuTI (41739/31/2020). K.D.J. acknowledges funding provided by the Deutsche Forschungsgemeinschaft (DFG, German Research Foundation) – SFB-Geschäftszeichen TRR142/3-2022 – project number 231447078. This work is also supported by ERC grant (LiNQs, 101042672).

**Author contributions**

A.C., M.G., and T.H. conceived the foundational ideas and designed the experiments. A.C. assisted with various steps of the nanofabrication process, performed materials and optical characterization, and wrote the first draft of the manuscript. T. H. performed all FDTD and Matlab simulations, and co-ordinated the entire project. R.H., under the supervision of M.G., carried out epitaxial growth of the samples. S.B. and H.W., under the supervision of T.N., carried out various nanofabrication tasks including metallization, dry-etching, and template-stripping. T.U. and H.K. assisted with template-stripping. H.R. performed EBL patterning under the supervision of P.K. T.K., C.S., and T.H. mapped the optical characteristics of QD-nanopillar arrays. E.S., L.H., P.K. and A.C., under the supervision of K.D.J., carried out optical characterization of single QDs. All authors contributed to data analysis and edited the manuscript.

**Competing interests**

The authors declare no competing interests.

# Supplementary Information:

**Mode resonance of Ag-coated nanopillars**

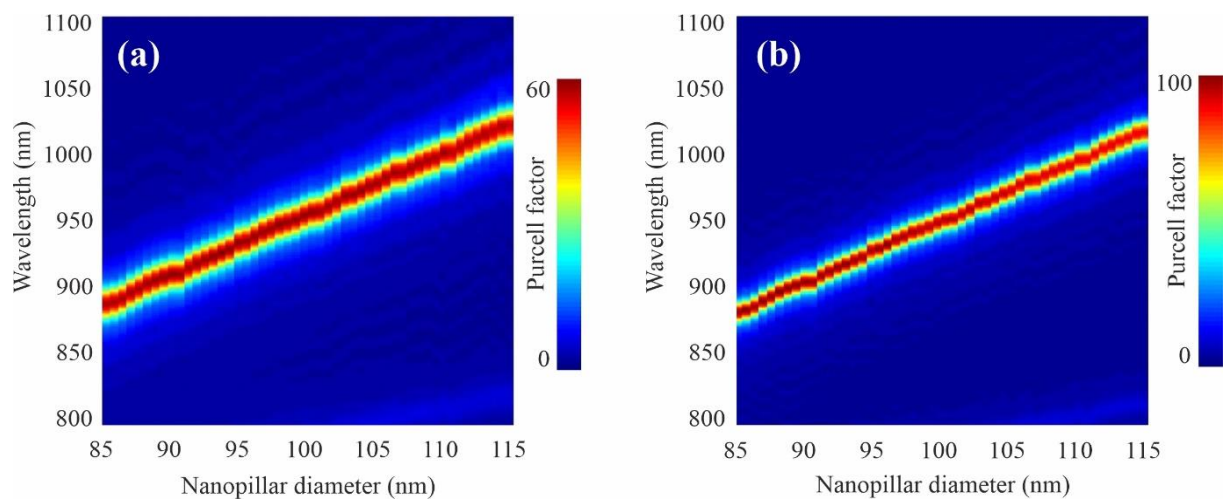

**Fig. S1 Simulation of cavity mode resonance.** Simulated mode resonance vs nanopillar diameter vs $F_P$ for the case when the nanopillar is surrounded by **a,** rough Ag and **b,** smooth Ag.

The cavity mode resonance wavelength red-shifts as a function of increasing nanopillar diameter, $d$. The mode resonance is evidently broader in the case of rough Ag surrounding the nanopillar. The simulated values of the Purcell factor, $F_P$, are shown as a function of $d$ and mode resonance wavelength in **Fig. S1**. The maximum value of estimated $F_P$ reaches up to 60 in the case of rough Ag surrounding the nanopillar (**Fig. S1a**). Whereas smooth Ag surrounding the nanopillar allows reaching values of $F_P$ up to 90 (**Fig. S1b**).

**Nanofabrication**

An atomic force microscopy (AFM) image of the self-assembled InAs QDs is shown in **Fig. S2a**. The density of the QDs is estimated to be $2\times10^{10}$ cm$^{-2}$. The yellow circle on the image is representative of the footprint of a nanopillar[100] overlapping with a single QD aligned to the center of the cavity. The photoluminescence (PL) emission peak of the QD ensemble under non-resonant excitation is at 925 nm at a temperature of 6 K (**Fig. S2b**). The full-width-at-half-maximum (FWHM) of the inhomogeneously broadened emission peak, estimated by fitting a Gaussian function, is 27 nm. **Figure. S2c** shows a schematic of the layer structure in the sample(s) which underwent the *flip-strip* process. **Figure. S2d** shows the scanning electron microscopy (SEM) image of the cross-section of a metallized GaAs nanopillar. The cross-section was obtained by focused ion beam etching. It must be noted that this image was captured during the early runs of processing; the physical characteristics of the nanopillar(s) and the Ag layer have since been optimized/improved. Nevertheless, this image provides evidence for a conformal and uniform coating of Ag along the interfaces of the nanopillar.

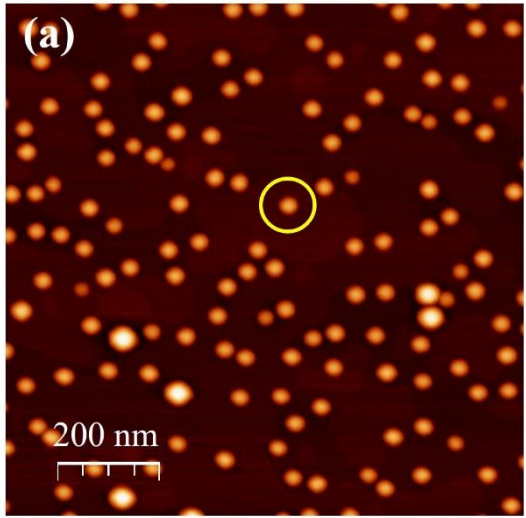
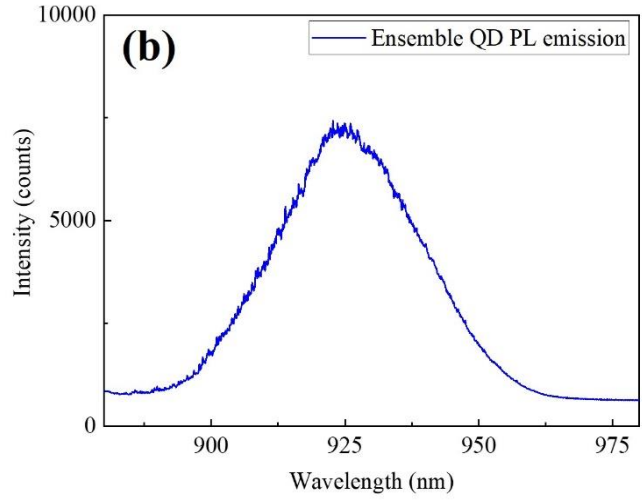
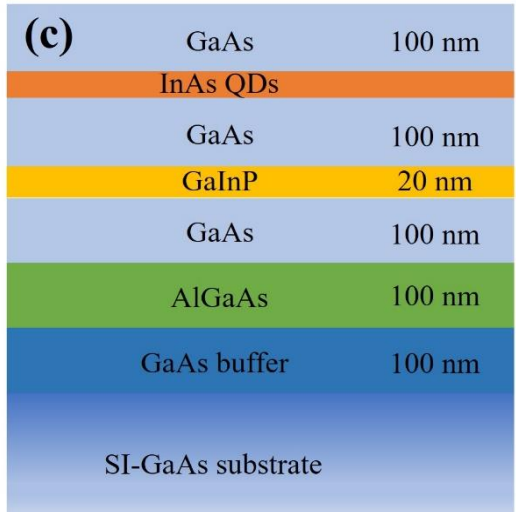
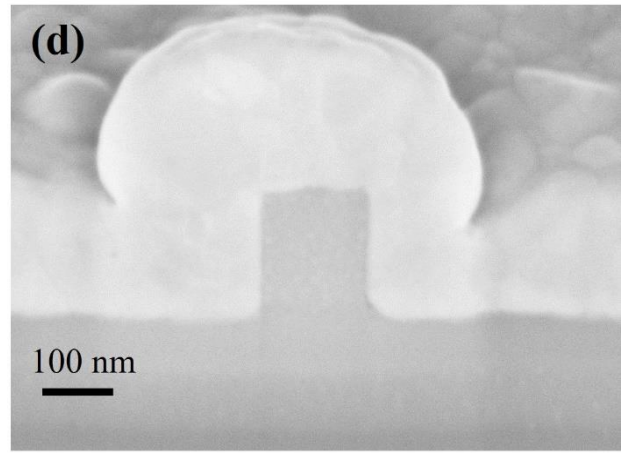

**Fig. S2 Nanofabrication. a,** AFM image of self-assembled InAs QDs on a GaAs matrix. The yellow circle represents the scenario when a single QD is aligned to the center of a nanopillar[100]. **b,** PL emission of the QD ensemble under non-resonant excitation at 6 K. **c,** Schematic representation of the layer structure of sample(s) utilized for optical experiments. **d,** SEM image of the cross-section of a metallized GaAs nanopillar.

## Time-resolved photoluminescence mapping

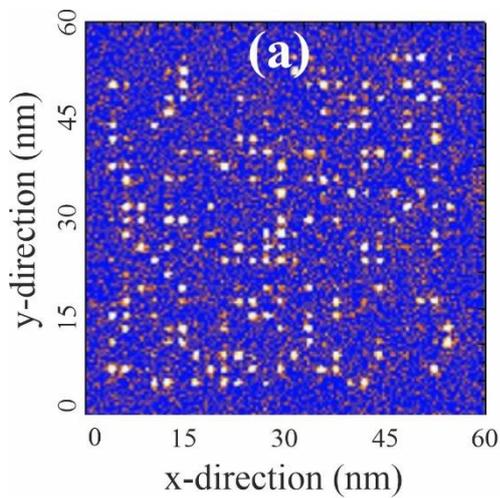
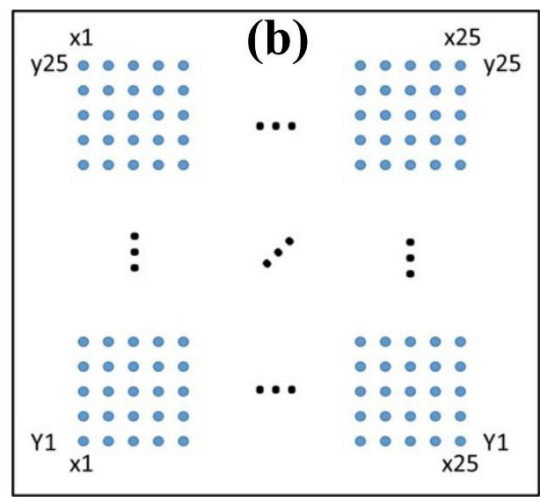

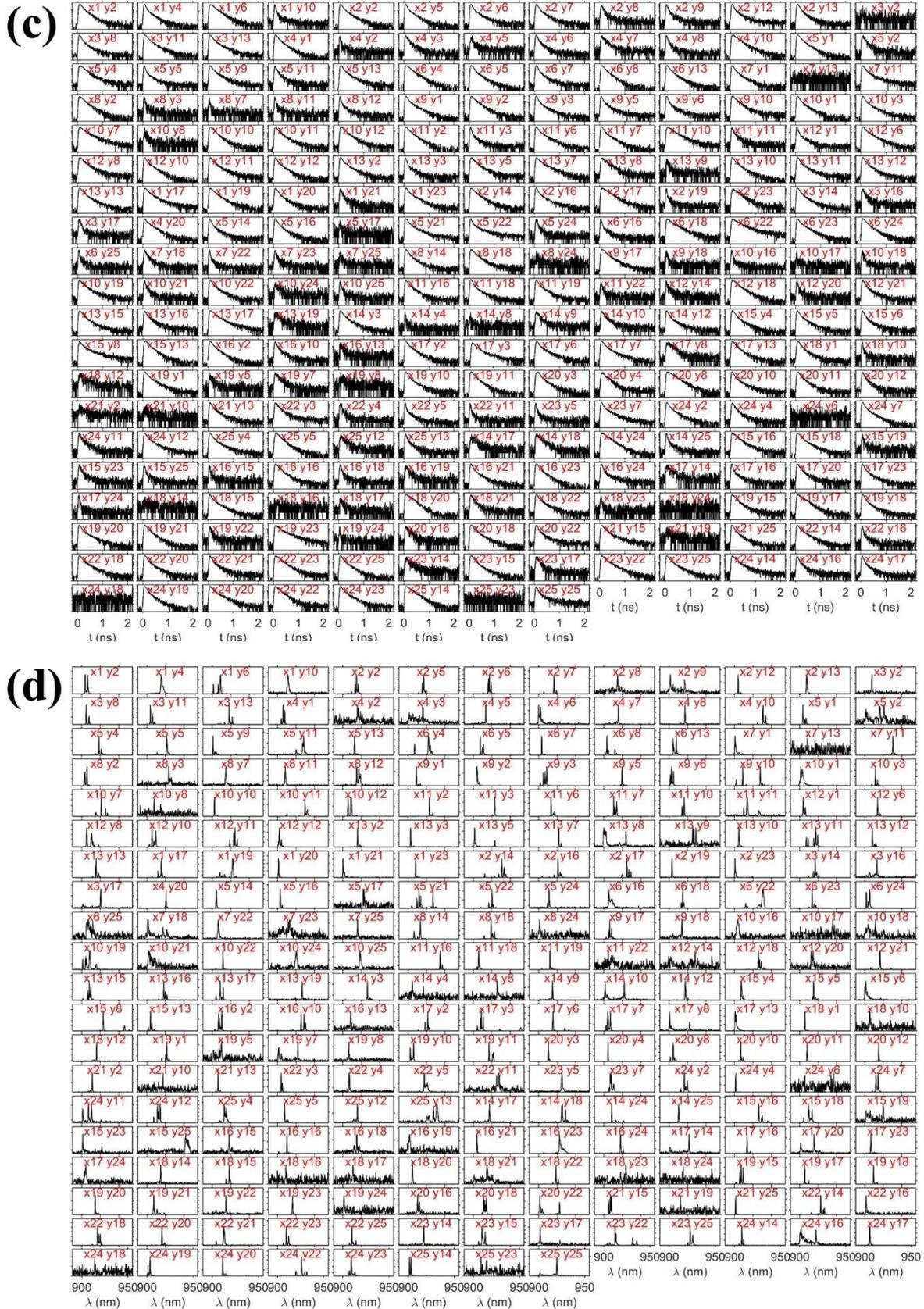

**Fig. S3 TRPL mapping. a,** PL intensity map of QD-nanopillar units in a single 25×25 array. **b,** Schematic representing the choice of initial coordinates when mapping QD-nanopillars in a 25×25 array. **c,** TRPL responses of individual QD-nanopillar units obtained by deterministically moving to targeted coordinates within a 25×25 array. **d,** PL spectra corresponding to the TRPL responses.

**Figure. S3a** shows a PL intensity map obtained from a 25×25 array of QD-nanopillar units by scanning the laser over the sample using a Galvo-scanner as described in the Methods section. Such a map was used as a guide to set the coordinates of each nanopillar location within an array. The choice of initial coordinates was sketched as shown in **Fig. S3b**. Next, the spectra collected from all nanopillar positions in the array were measured automatically as described in the Methods section. By manually inspecting the recorded spectra, an intensity threshold was used for filtering the optically active QD-nanopillar units within each array. Subsequently, the Galvo-scanner was programmed to move deterministically to the location(s) of the filtered QD-nanopillar units based on the coordinate system. The time-resolved PL (TRPL) response was collected from all optically active QD-nanopillar units within an array (**Fig. S3c**). The decay curves were normalized to their maximum and the FWHM was then extracted from each decay response and plotted in a 25×25 pixel map (as shown in **Figs. 4a–4g**). Ultimately, these TRPL maps allowed us to study decay statistics from QD-nanopillar units as a function of $d$. For the sake of completeness in reporting the analyzed data, the PL spectra of all QD-nanopillar units corresponding to the decay curves in **Fig. S3c** are shown in **Fig. S3d**.

**FWHM vs single-exponential fit for estimating $\tau$**

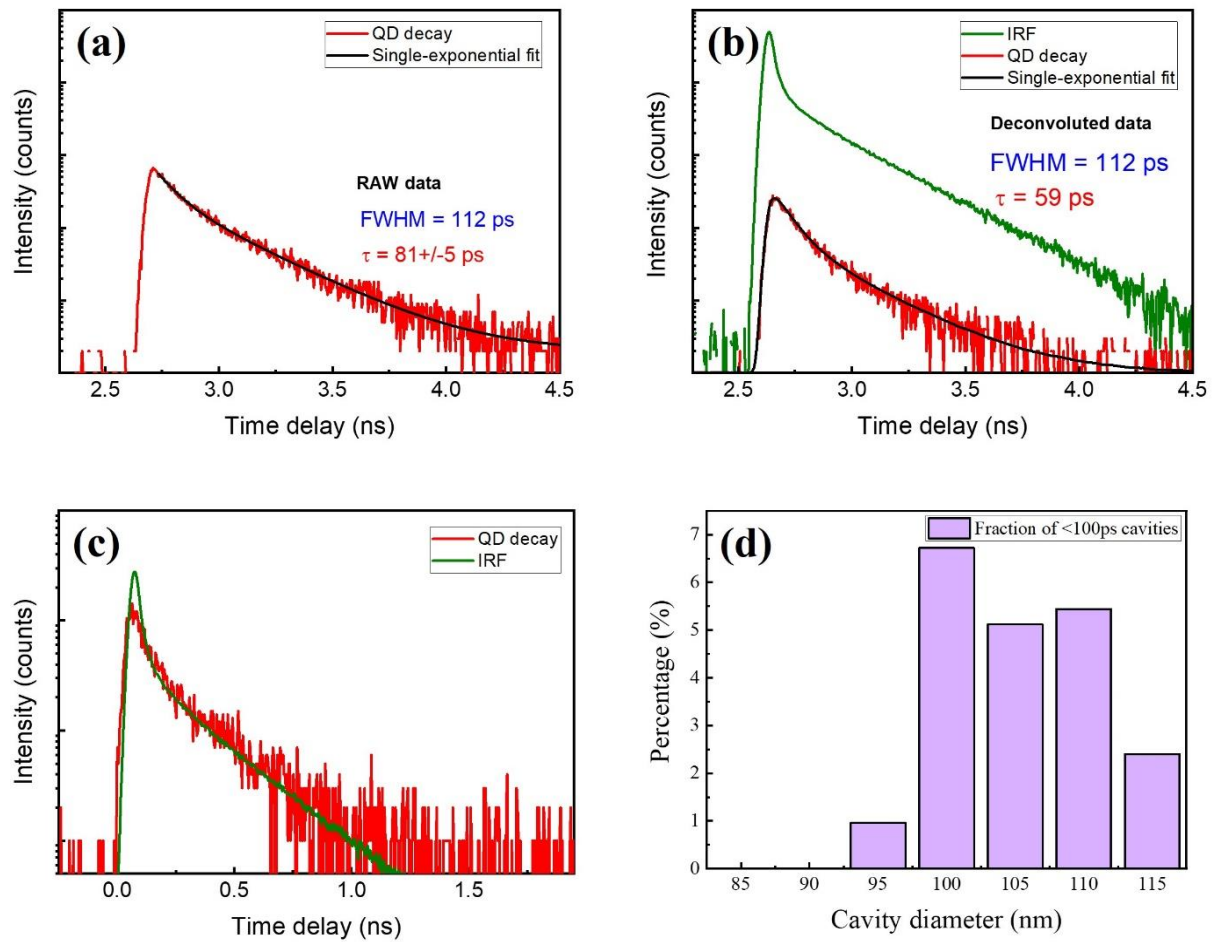

**Fig. S4 Estimating $\tau$ using FWHM and single-exponential fit. a,** Raw decay curve collected from a QD. **b,** The decay curve after deconvolution with the IRF. **c,** Raw decay curve from a QD whose emission lifetime is comparable to the IRF. **d,** Fraction of optically active QD-nanopillar units exhibiting FWHM < 100 ps.

Here, it is necessary to make the distinction between estimating $\tau$ from the FWHM of a decay curve and from fitting it with a single-exponential function. The FWHM is consistently larger than the value of $\tau$ obtained through a single-exponential fit of the raw decay curve (**Fig. S4a**). However, the FWHM is considerably less error-prone when obtaining useful information from data close to the resolution limit of the measurement system. Thus, it is more suitable for this purpose where it is important to extract robust single-pixel information reliably from a large dataset containing raw decay responses of varying intensities and linewidths (and often accompanied by a considerably noisy background). Fitting a single-exponential function to a decay curve deconvoluted with the instrument response function (IRF) might provide a more accurate estimate of $\tau$ (**Fig. S4b**). However, it is not possible to deconvolute decay curves with FWHM <100 ps due to the system-defined timing resolution in the setup used for TRPL mapping (**Fig. S4c**).

In this regard, QDs with the shortest decay lifetimes end up being represented by the system-defined resolution-limited FWHM values of ~100 ps (i.e, 50 ps detector resolution) in the pixel maps shown in **Figs. 4a–4g**. By calculating the total number of system-defined resolution-limited pixels, i.e, FWHM <100 ps, we can determine the cavity dimension that is most suitable for providing Purcell enhancement. Evidently, the fraction of optically active QDs exhibiting FWHM <100 ps is highest in the nanopillars$|^{100}$ region (**Fig. S4d**). This result supports the conclusion we arrived at by judging the most suitable cavity dimension on the basis of $\bar{\tau}_{FWHM}$ (**Fig. 4h**). It should be noted that the estimated probability of having a QD located within 10 nm from the center of the cavity is 6.3% for uniformly distributed QDs with a density of $2\times10^{10}$ cm$^{-2}$, which matches remarkably well with the observed percentage of QDs with FWHM <100 ps.

**Effects of radial displacement in the position of a QD within a nanopillar**

Simulations reveal that the radial displacement of a QD from the center of the nanopillar reduces the $F_P$ (**Fig. S5a**) and introduces an anisotropy in its emission polarization (**Fig. S5b**). In the case of a radially oriented dipole, the estimated $F_P$ is relatively invariant for a radial displacement up to ~20 nm from the axis of the nanopillar. Beyond that, the estimated $F_P$ decreases from 60 to 50 for a radial displacement up to 45 nm. The Purcell enhancement experienced by a tangentially oriented dipole, however, is more strongly affected by radial displacement from the axis of the nanopillar. In this case, the $F_P$ is seen to decrease from 60 to 20 as the radial displacement increases from 0 nm to 45 nm. Considering their finite lateral dimensions, QDs displaced by more than 30 nm are expected to be optically inactive because they too close to the sidewall(s) of an etched nanopillar.

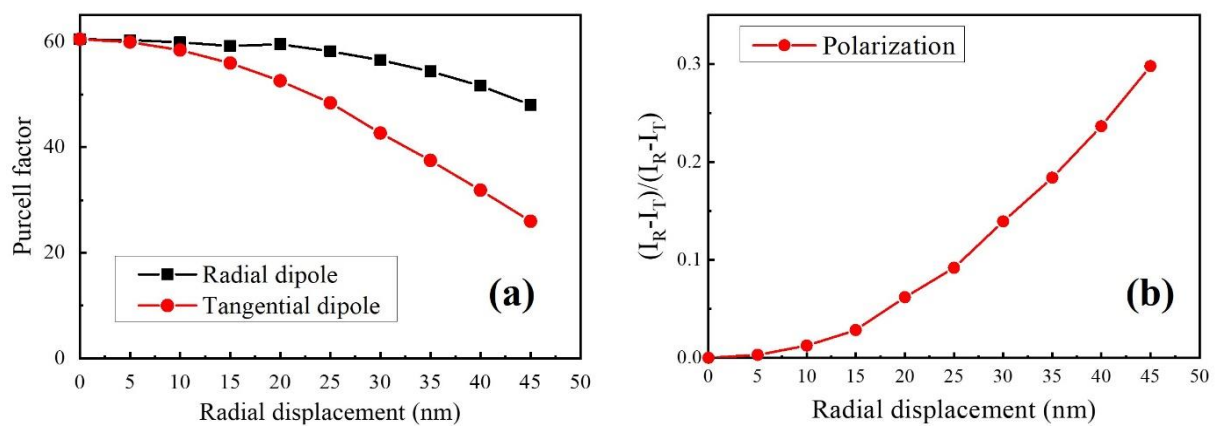

**Fig. S5 Effects of radial displacement of QD within a nanopillar. a,** The estimated $F_P$ as a function of radial displacement for radially and tangentially oriented dipole emitters placed within a nanopillar. **b,** The effects of radial displacement of the dipole on the polarization of the emitted photons collected from the nanopillar. The induced phase-shift increases as a function of the displacement.

It must be noted that these simulations are made for an emitter in a perfectly symmetric cylindrical cavity. The polarization anisotropy of emission is seen to increase as the radial displacement increases from 0 nm to 45 nm (**Fig. S5b**). Importantly, this component of polarization anisotropy exists in addition to the anisotropy caused by the ellipticity of the fabricated cavities (**Figs. 4h** and **5c**). Therefore, the polarization response arising from the ellipticity **(Fig. 5)** was assessed only for the fastest QDs from respective cavity arrays, i.e. the QDs located at the center of the nanopillar where the Purcell enhancement is the strongest and the displacement plays no role.

## Effects of ellipticity of the nanopillar

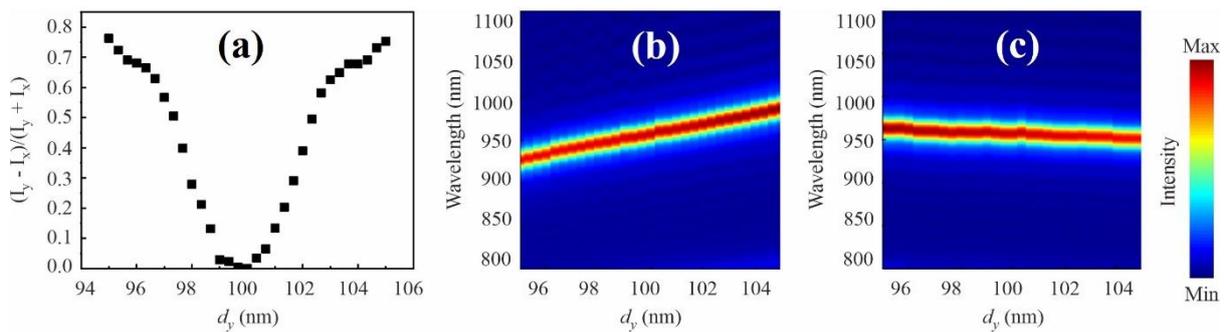

**Fig. S6 Effects of ellipticity of the nanopillar. a,** Simulation of polarization anisotropy in an elliptical nanopillar. The anisotropy is seen to increase as a function of increasing ellipticity/asymmetry in the nanopillar diameter. **b,** Intensity sweeps made for x-polarization as a function of increasing asymmetry. **c,** Intensity sweeps made for y-polarization as a function of increasing asymmetry.

Simulations were carried out for a dipole situated at the center of an elliptical nanopillar where the dimension in x-direction, $d_x$, is maintained constant at 100 nm while $d_y$ is changed. The anisotropy in emission intensity along the x- and y-polarizations is seen to increase as a function of increasing $d_y$ (**Fig. S6a**). A closer look at the dependency of intensities along x- and y-polarizations as a function of increasing asymmetry reveals that the dipole is more sensitive to the dimension perpendicular to the polarization (**Figs. S6b** and **S6c**).

**Other notable Purcell-enhanced QDs**

The TRPL responses of other notable single QDs embedded within nanopillars and measured under pulsed incoherent p-shell excitation are shown in **Fig. S7**. The $g^{(2)}(0)$ data shown in **Fig. 6b** corresponds to the decay lifetime of QDs shown in **Fig. S7**.

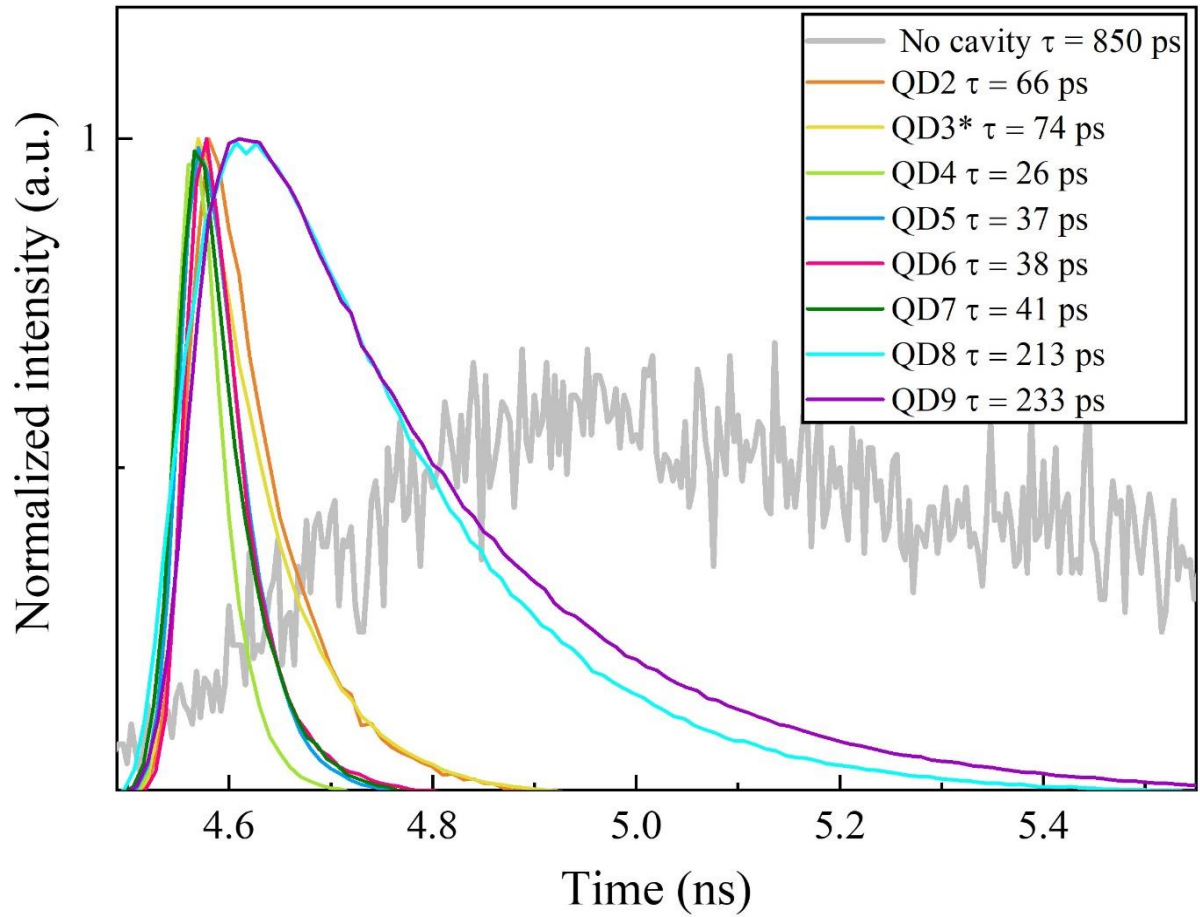

**Fig. S7 TRPL response from other notable Purcell-enhanced single QDs.** The TRPL response collected from single QDs under incoherent p-shell excitation. The gray curve represents the decay response of a QD ensemble obtained via exciting their wetting layer.

**Linewidth of the QD emission peaks**

In the ideal case, the linewidth of an emission peak originating from a radiative transition within the QD is determined solely by its radiative lifetime. This is termed as the Fourier-limited linewidth and the shape of the emission peak in this case is purely Lorentzian. However, in practice, the experimentally measured peak best fits the shape of a Voigt function. This is because the Lorentzian emission line is convoluted with the spectral response function of the spectrometer, which can be considered as Gaussian for small slit sizes[1]. So, the Gaussian broadening component has to be separated from the measured peak in order to determine the "true" linewidth of the emission from the QD.

As an example, we consider here the case of QD1 which is characterized with a $\tau = 22.59$ ps and exhibits an emission peak linewidth of 63.4 µeV. This is determined from the Voigt function fitted to the experimental data, as shown in **Fig. S8**. The Fourier-limited linewidth of the emission peak with $\tau = 22.59$ ps would be $\Delta E_{Fourier} = 29.14$ µeV. The ultra-fast lifetime of strongly Purcell-enhanced QD1 allows for applying a deconvolution method similar to what is used in interferometric measurement of linewidths[2]. We determine the true linewidth of the QD1 emission peak by setting a resolution-limited

Gaussian component of 40 μeV (determined from our measurement setup) and estimating the linewidth of the resulting deconvoluted Lorentzian peak. This is found to be $\mathit{\Delta E_{Deconv}}$ = 36.6 μeV, as shown in **Fig. S8**. The deconvoluted linewidth of QD1 is larger than the Fourier limit by a factor of only 1.25, which suggests that the radiative transition is negligibly influenced by sources of spectral diffusion.

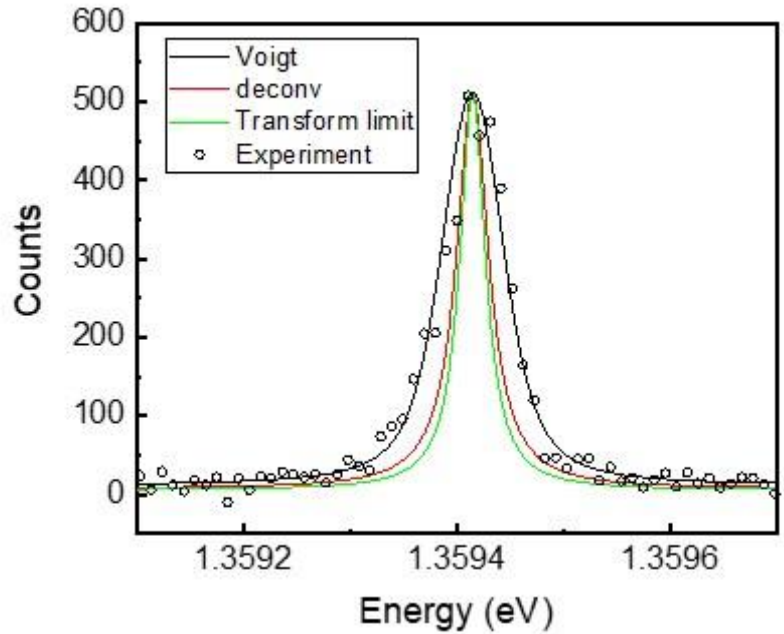

**Fig. S8 Linewidth of the QD emission peak.** The experimentally measured emission peak of QD1 is fitted with a Voigt function. The Fourier-limited emission peak and the deconvoluted Lorentzian emission peak are both shown in the plot.